# Self-Adaptive Active Damping Method for Stability Enhancement of Systems With Black-Box Inverters Considering Operating Points

Yang Li, Xiangyang Wu, Zhikang Shuai, *Senior Member*, *IEEE*, Junbin Fang, Lili He, *Member*, *IEEE*, Yi Lei, and Z. John Shen, *Fellow*, *IEEE*

*Abstract*—Due to the black-box nature of inverters and the wide variation range of operating points, it is challenging to on-line predict and adaptively enhance the stability of inverter-based systems. To solve this problem, this paper provides a feasible self-adaptive active damping method to eliminate potential small-signal instability of systems with black-box inverters under multiple operating points. First, the framework that includes grid impedance estimation, inverters' admittance identification, and self-adaptive strategy is presented. Second, a widely-applicable and engineering-friendly method for inductive-resistive grid impedance estimation is studied, in which a frequency-integral-based dq-axis aligning method is presented to avoid the inaccuracy resulting from the disturbance $\Delta\theta$. Then, to make the system have a sufficient stable margin under different operating points, a self-adaptive active damper (SAD) as well as its control strategy with lag compensator modification is proposed, in which the SAD's damping compensation mechanism for the system's stability enhancement is investigated and revealed. Finally, the mapping between system's parameter variations and SAD's parameters is established based on the artificial neural network (ANN) technique, serving as a computationally light model surrogate that is favorable for on-line parameter-tuning for SAD to compensate the system's damping according to operating points. The effectiveness of the proposed method is verified by simulations in PSACD/EMTDC and experiments in RT-Lab platforms.

*Index Terms*—Inverter-based systems, small-signal stability, self-adaptive active damper (SAD), black-box inverter.

## I. INTRODUCTION

POWER electronic inverters have been widely used as the interface devices for delivering the power generated by wind/solar to the grid. However, with the large-scale integration of inverters, the instability issues caused by the interactions between inverter's control loops and grid impedance seriously threaten the safety and reliability of power systems [1]-[3]. Considering the black-box nature of inverters (their specific parameters could not be obtained because of commercial confidentiality) and the wide variations range of operating points, on-line prediction and elimination of the instability in inverter-based systems are of great significance [4].

Researches have shown that weak AC grids with low short circuit ratio likely lead to small-signal instability issues in converter-based systems [3], [5]. Accurately estimating the gird impedance is important to analyze and stabilize the systems with inverters. Ref. [6] summarizes the typical ways to estimate the grid impedance and presents a method for estimating the gird impedance of inductive-resistive power networks by using an extended Kalman filter. This method does not interfere with the system's normal operation but has the drawback of relatively low estimation accuracy [7]. To overcome this drawback, the grid impedance estimation method based on the grid-tied inverter's output power variations is studied in [7]-[9], where active power and reactive power perturbations are required. To simplify the estimation process and make the inverter have a higher power factor, [10] proposed a grid impedance estimation method that only needs active power perturbations. However, this method is only suitable for the inverters with a $P$-$\omega$ droop control loop. Therefore, a grid impedance estimation method with a wider range of applications should be explored.

The impedance-based stability criteria and the impedance measurement technique are commonly and powerfully used to analyze the stability of systems with black-box inverters [11]-[13]. The impedance-based stability criteria assessing the stability of inverter-based systems rely on the port impedances/admittances of devices and network components rather than the complicated state-space model that requires the specific parameters of the whole system. Fortunately, the port impedances or admittances of black-box inverters at a specific operating point can be obtained by the frequency scanning based-impedance measurement technique. To acquire the inverter's accurate and comprehensive model that considers variable operating points, artificial neural network (ANN)-based methods have been developed and recently of great interest [14]-[16]. Ref. [14] applies a general framework for applying ANN to identify the impedance models of inverters considering the operating-point variation. Ref. [15] presents a machine-learning framework capable of characterizing inverter impedance patterns to model black-box inverter-based systems. The stability region of the black-box grid-tied inverters is estimated by a deep neural network-based method [16], which can effectively estimate the small-signal stability of the wide variation of operating points caused by large variations of renewable energy. Overall, ANN-based technique has promising application prospects in the stability assessment of power systems with black-box inverters.

The purpose of stability assessment or stability region identification is primarily to stabilize the systems. Ref. [17] proposes an ANN-based pole-tracking method to establish the mapping between the control parameters and the system's closed-loop poles for on-line stabilization control. Another ANN-based adaptive multi-parameter-tuning method for on-line stabilization control of grid-tied converters is proposed in [18], which can achieve the converter's control adaptability to the varying grid conditions. Ref. [19] presents an adaptive virtual-inductance feedforward scheme to enhance the stability of inverters in weak grids, in which the virtual inductance is updated in real-time according to an adaptation law to ensure the system's stability. However, these techniques proposed in [17]-[19] involve control parameter tuning or control strategy modification of the inverters, which means these inverters must be white-box ones, thus are not applicable for the systems with black-box inverters that nearly have zero controllability.

The utilization of additional devices for active damping of inverter-based systems has been studied in recent years. Ref. [20] adopts an extra active damper for the stability enhancement of microgrids with unknown-parameter inverters, where installing position and parameter design of the damper are discussed, but the impact of the variations of operating points is not considered. Ref. [21] presents an adaptive active damper to cope with the instability of the gird-tied inverter caused by variations of grid impedance. This damper's work state is determined by whether the harmonic voltage (caused by instability) is higher than a certain threshold or not. It means that the damper starts working after the instability phenomenon arose, which perhaps is unfavorable for the safety and reliability of the systems. Undoubtedly, a damper that can eliminate the risk of potential instability issues—instead of roughly suppressing the distortion caused by instability—will be the most beneficial to the inverter-based systems.

To eliminate the risk of potential instability for the systems with black-box inverters, this paper proposes a self-adaptive active damping method based on the ANN technique, which enables the system to have a sufficient stable margin when its parameters (including inverters' output power and grid impedance) vary. The main work and contributions of this paper are summarized as follows:

1) An adaptive active damping framework, including on-line grid impedance estimation, inverter admittance identification, and self-adaptive active damping strategy, is proposed to eliminate the risk of potential instabilities in black-box inverter-based systems, which can ensure the system has a sufficient stability margin under a wide range of operating points.
2) A frequency-integral-based dq-axis aligning method for grid impedance estimation is presented, which only requires applying reactive power disturbance and is therefore engineering-friendly and application-widely to different types of converter devices.
3) The control strategy with a lag compensator (LAC) for the self-adaptive active damper (SAD) is presented, and the SAD's damping compensation mechanism for the system's stability enhancement is also revealed. Based on this, the parameters of SAD can be properly designed under different operating points to ensure the system has a sufficient stability margin.

The remainder of this paper is organized as follows, with an outline in Fig. 1. Section II outlines the proposed method's framework and details the impedance identification of an inverter-based system. Section III elaborates on the self-adaptive strategy for the active damper aimed at enhancing the system's stable margin. Experimental validations are presented in Section IV, followed by the conclusions in Section V.

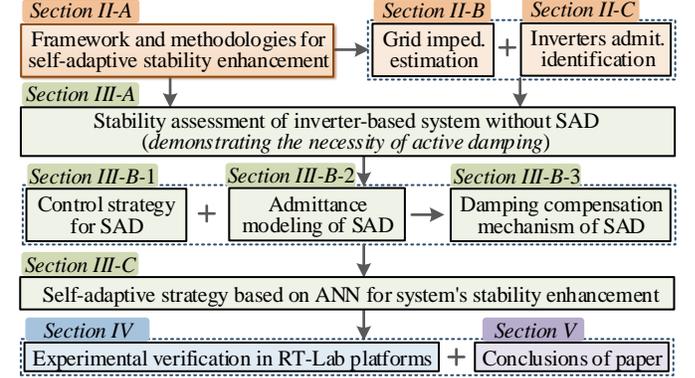

Fig. 1. The outline of this paper.

## II. FRAMEWORK OF PROPOSED METHOD AND IMPEDANCE IDENTIFICATION OF INVERTER-BASED SYSTEM

### A. Framework and Methodologies of Self-Adaptive Active Damping for Paralleled Grid-Following Inverters

Fig. 2 depicts the framework and methodologies of the proposed method. A typical three-phase multiple inverter-based system, where the grid-following (GFL) inverters ($inv.1$ - $inv.n$) are controlled as current sources and deliver the power generated by wind or solar to the grid. In a weak grid, the interaction between the control loops of inverters and passive components of the system could lead to instability issues. Even if the parameters of inverters have been designed properly, alterations of the operating conditions of inverters and the grid impedance could deteriorate the system's stability [3], [20]. To eliminate the risk of potential instability issues, this paper proposes a SAD for stability enhancement, which can adaptively compensate the system's damping according to inverters' operating conditions and grid impedance.

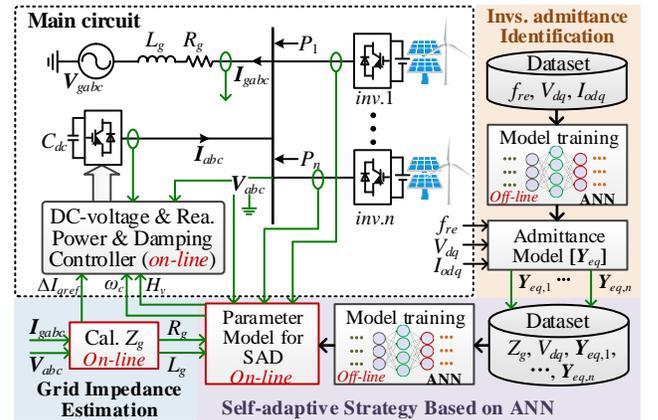

Fig. 2. Framework and methodologies of self-adaptive active damping for multiple paralleled grid-following inverter-based system with SAD.

As illustrated in Fig. 2, the framework consists of three parts: inverters' admittance identification, self-adaptive active damping strategy, and grid impedance estimation. The first two parts involve the ANN technique, and their ultimate objective is to generate the parameter model, which serves as a computationally light model surrogate that is favorable for on-line parameter-tuning for SAD according to operating points. The last part calculates the grid impedance that combined with the inverters' output currents and point of common coupling point (PCC) voltage to adaptively adjust the SAD parameters (i.e., $\omega_c$ and $H_v$) for SAD on-line according to the operating points.

The proposed SAD is supposed to be connected to PCC, and its topology and control block diagram can be found in Fig. 3. The SAD adopts a voltage-current dual closed-loop control strategy, in which the voltage control loop put on the d-axis is used for dc-link voltage regulation, and the reactive current $I_q$ is controlled for reactive power compensation (if necessary) and on-line grid impedance estimation, which will be discussed in next part. In addition, a self-adaptive damping control strategy is employed, which will be discussed in Section III.

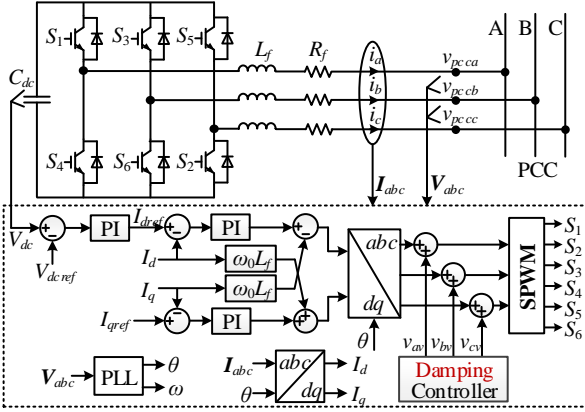

Fig. 3. Topology and control block diagram of the proposed SAD.

### B. On-line Grid Impedance Estimation by Frequency-Integral-Based dq-Axis Aligning Method

In the steady state, the relationship of the grid voltage $V_g$ and impedance $Z_g$, PCC voltage $V$, and grid current $I_g$ can be expressed as

$$V = Z_g I_g + V_g \quad (1)$$

where $V_g=[V_{gd}, V_{gq}]^T$, $V=[V_d, V_q]^T$, $I_g=[I_{gd}, I_{gq}]^T$, $Z_g=[sL_g+R_g, -\omega_0 L_g; \omega_0 L_g, sL_g+R_g]$, $\omega_0$ is the fundamental frequency of the grid, and $R_g$ and $L_g$ denote the equivalent resistance and inductance of grid impedance. All these variables are expressed in dq axis.

Based on (1), considering any two steady operating points of the system, we have

$$V_1 = Z_{g1} I_{g1} + V_{g1} \quad (2)$$
$$V_2 = Z_{g2} I_{g2} + V_{g2} \quad (3)$$

where the subscripts "1" and "2" denote the related variables picked at operating point 1 and operating point 2, respectively.

In a short time-scale, e.g., within 2 s, the grid voltage is supposed to be constant, hence $V_{g1}=V_{g2}$. Then subtracting (2) from (3) yields

$$\begin{bmatrix} \Delta V_d \\ \Delta V_q \end{bmatrix} = \begin{bmatrix} sL_g + R_g & -\omega_0 L_g \\ \omega_0 L_g & sL_g + R_g \end{bmatrix} \begin{bmatrix} \Delta I_{gd} \\ \Delta I_{gq} \end{bmatrix} \quad (4)$$

where $[\Delta V_d, \Delta V_q]^T = V_2 - V_1$, and $[\Delta I_{gd}, \Delta I_{gq}]^T = I_{g2} - I_{g1}$.

For the DC components of voltages and currents in dq-axis (i.e., the fundamental components of system's voltages and currents in abc-axis), $s=j0$, then (4) can be rewritten as

$$\begin{bmatrix} R_g \\ L_g \end{bmatrix} = \begin{bmatrix} \Delta I_{gd} & -\omega_0 \Delta I_{gq} \\ \Delta I_{gq} & \omega_0 \Delta I_{gd} \end{bmatrix}^{-1} \begin{bmatrix} \Delta V_d \\ \Delta V_q \end{bmatrix} \quad (5)$$

Eq. (5) means that $R_g$ and $L_g$ could be on-line calculated when $\Delta V_d$, $\Delta V_q$, $\Delta I_{gd}$, and $\Delta I_{gq}$ can be obtained. This is achievable when the active power $P_g$ (referring to $I_{gd}$) and/or reactive power $Q_g$ (referring to $I_{gq}$) injecting into the grid are different at the operating points 1 and 2. The previous work in [7]-[9] achieved the on-line grid impedance calculation by adjusting both $P_g$ and $Q_g$, respectively. This paper proposes a more practical procedure to calculate $R_g$ and $L_g$ on-line by only adjusting $Q_g$, while the variation of $P_g$ is dispensable ($P_g$ depends on the output active power of the grid-tied inverters).

The adjustment of $Q_g$ can be fulfilled by altering a $\Delta I_{qref}$ (addition or subtraction) to $I_{qref}$ of SAD. As illustrated in Fig. 4(a), a $\Delta I_{qref}$ is added to $I_{qref}$ at $t_1$, meanwhile the DC components of PCC voltage is recorded, which is marked as $V_d(t_1)$ and $V_q(t_1)$; after a short time, $I_{qref}$ restores to the original value at $t_2$ (ensuring that the system has reached a steady state), and the DC components of PCC voltage is recorded again.

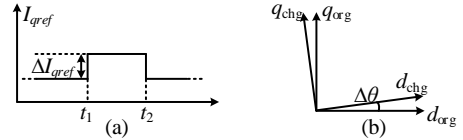

Fig. 4. Reactive power disturbance applied to the grid: (a) the procedure for the variation of $I_{qref}$, and (b) dq-axis before and after changing $I_{qref}$.

Since changing $I_{qref}$ will introduce a disturbance $\Delta\theta$ to the dq steady axis, as depicted in Fig. 4(b), the DC components of PCC voltage picked in the dq axis after changing $I_{qref}$ will lead to inaccurate calculation of $R_g$ and $L_g$. Researchers use the variations of both $P_g$ and $Q_g$ to avoid the inaccuracy resulting from $\Delta\theta$ in the dq axis and obtain accurate grid impedance in [7]-[9]. To simplify the process of the grid impedance estimation, Ref. [10] uses only the variation of $P_g$ to obtain accurate grid impedance; however, this method requires adjusting the active power reference of the P-$\omega$ droop control loop of the grid-forming (GFM) inverter, which means its application is limited and not applicable to different types of converter devices.

To solve the above problem, this paper proposes a frequency-integral-based dq-axis aligning method, which can pick $V_d(t_2)$ and $V_q(t_2)$ in the same dq axis before changing $I_{qref}$, then accurate $R_g$ and $L_g$ can be calculated. As illustrated in Fig. 5, the phase and frequency outputs from the phase-locked loop (PLL) are supposed to be also recorded at $t_1$, marked as $\theta(t_1)$ and $\omega(t_1)$, then $\theta(t_1)$ plus the integral of $\omega(t_1)$ to obtain $\theta(t_2)$; subsequently, $\theta(t_2)$ is used to transfer $V_{abc}$ from abc-axis to dq-axis to obtain $V_d(t_2)$, $V_q(t_2)$. The DC components of gird current $I_{gd}(t_1)$, $I_{gq}(t_1)$, $I_{gd}(t_2)$, and $I_{gq}(t_2)$ can also be obtained in a similar way.

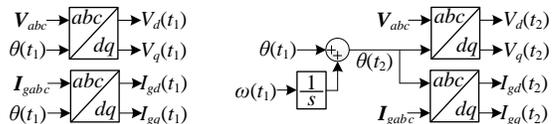

Fig. 5. Obtaining PCC voltage and grid current before and after changing $I_{qref}$ in the same dq axis.

As a matter of fact, the grid frequency $\omega_g$ can be regarded as a constant in a short time-scale, so that $\theta(t_2)$ can be obtained in an easier way, i.e., $\theta(t_2)=\theta(t_1)+\omega_g(t_2-t_1)$. By means of this way, $V_d(t_2)$, $V_q(t_2)$, $V_d(t_1)$, $V_q(t_1)$, $I_{gd}(t_2)$, $I_{gq}(t_2)$, $I_{gd}(t_1)$, and $I_{gq}(t_1)$ will be in the same dq-axis, then the error of the calculation of $R_g$ and $L_g$ led by the disturbance $\Delta\theta$ can be eliminated.

The traditional technique based on injecting harmonics to estimate grid impedance has the disadvantage of degrading the power quality of the system. Moreover, such type of techniques is prone to be affected by background harmonic noise interference. Since that the proposed method uses the DC components of the system's voltages and currents to calculate $R_g$ and $L_g$ on-line, the high-frequency noise interference in the background of the grid can be easily eliminated by a low-pass filter (LPF). Compared with the conventional techniques that require the variations of both $P_g$ and $Q_g$ in [7]-[9], the proposed grid impedance estimation method in this paper only needs to adjust the reactive power, which would never affect the maximum power point tracking of the inverters. And compared with the active power variation-based technique for GFM inverters in [10], the method in this paper has wider applicability, not only applicable to GFL and GFM inverters, but also to the SAD proposed in this paper. The comparison of grid impedance estimation methods is summarized in Table I.

TABLE I
COMPARISON OF GRID IMPEDANCE ESTIMATION METHODS

| Methods | Required data | Adaptation |
| --- | --- | --- |
| [7]-[9] | both $P_g$ and $Q_g$ | GFLs with power control |
| [10] | $P_g$ | GFMs with P-$\omega$ droop control loop |
| This paper | $Q_g$ ($P_g$ is dispensable) | Both GFL, GFM, and SAD |

To verify the proposed grid impedance calculation method, the model of an inverter-based system with SAD is simulated in PSCAD/EMTDC, where the grid impedance is set as: $L_g$=3 mH and $R_g$=0.15 Ω. Fig. 6 demonstrates the simulation results of PCC voltage and grid current before and after power varying in the same dq axis. In the simulation shown in Fig. 6 (a), $\Delta I_{qref}$=40 A is added to $I_{qref}$ of SAD at $t_1$=2 s, and removed at $t_2$=3.5 s. It can get PCC voltage and grid current $[V_d(t_1), V_q(t_1)]^T$=[313.68 V, 0 V]$^T$, $[I_{gd}(t_1), I_{gq}(t_1)]^T$=[50 A, 0 A]$^T$, $[V_d(t_2), V_q(t_2)]^T$=[275.82 V, 5.20 V]$^T$, $[I_{gd}(t_2), I_{gq}(t_2)]^T$=[49.15 A, 40.90 A]$^T$ (note that the non-zero $I_{gd}$ is generated by GFL inverters rather than SAD), where first-order low-pass filters are used to extract these DC components. Then it can calculate the grid impedance as: $L_g$=2.937 mH and $R_g$=0.146 Ω.

In the simulation results shown in Fig. 6 (b), except the changing $I_{qref}$ of SAD, $I_{gd}$ generated by GFL inverters also changes (imitating inverter's active power variation), then one can get PCC voltage and grid current $[V_d(t_1), V_q(t_1)]^T$=[313.70 V, 0 V]$^T$, $[I_{gd}(t_1), I_{gq}(t_1)]^T$=[50 A, 0 A]$^T$, $[V_d(t_2), V_q(t_2)]^T$=[275.65 V, 30.51 V]$^T$, $[I_{gd}(t_2), I_{gq}(t_2)]^T$=[74.93 A, 48.68 A]$^T$. Then one can calculate the grid impedance as: $L_g$=2.936 mH and $R_g$=0.154 Ω. This means that the proposed method is also suitable for the scenario of both reactive and active power variations. The simulation results verify the effectiveness of the proposed grid impedance calculation method.

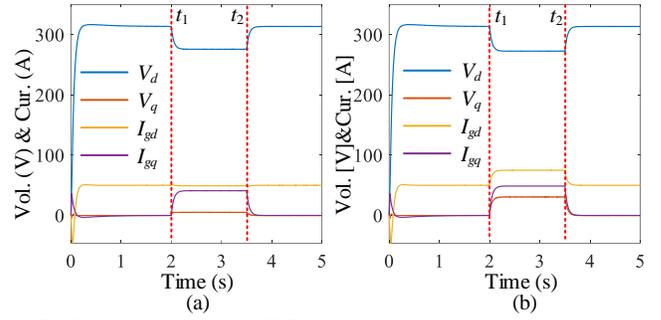

Fig. 6. Simulation results of PCC voltage and grid current before and after power varying in the same dq axis: (a) only reactive power varies, and (b) both reactive and active power vary.

### C. On-line Admittance Identification of Inverters

The inverter's output impedance/admittance is determined by its control strategy and parameters as well as operating points. Considering that the control strategy or parameters of commercial inverters are unavailable because of industry secrets and the grid-tied inverters are accompanied by a wide range of operating points, the ANN-based model generation technique has been proposed for the on-line identification of admittances of inverters [2].

Fig. 7 illustrates the structure diagram of the ANN, which consists of an input layer, a hidden layer, and an output layer. The input layer includes the frequency $f_{re}$, PCC voltage ($V_d$, $V_q$), and the inverter's output current ($I_{od}$, $I_{oq}$) in dq frame; while the output layer includes the real and imaginary parts, respectively, of the inverter's admittance in dq frame. The number of hidden layers and the number of neurons in each layer need to be selected based on experience. In this paper, 1 hidden layer and 10 neurons in the layer are used in the training of the ANN model, where the activation function of sigmoid is selected. Besides, the back-propagation algorithm is used to train the ANN model, where the typical mean squared error (MSE) is calculated to evaluate the loss function between the ANN output and measured impedance data features [14], [15]. After the training, the generated ANN model needs to be tested by the coefficient of determination ($R^2$).

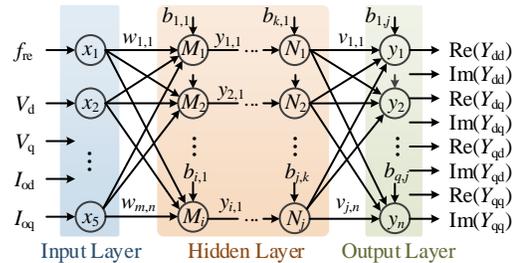

Fig. 7. Structure diagram of the ANN.

The dq impedance measurement technique is a useful tool to generate the source data for the training of the ANN model [14]. Firstly, the dataset that consists of operating points and the corresponding admittance of the inverter is established by the impedance measurement technique. Secondly, the dataset is randomly divided into three portions, namely 70% of the dataset is fed into the ANN for training, 15% and 15% are used for validation and testing, respectively [15]. Eventually, the on-line admittance identification of inverters can be achieved by the well-trained ANN model when the operating points of inverters (i.e., $V_d$, $V_q$, $I_{od}$, $I_{oq}$) are on-line measured and fed into the model.

Fig. 8 shows the error analysis of the model of admittance identification. The dataset of a GFL inverter (whose parameters can be found in Table I in Section III) is established by the impedance measurement technique based on the simulations in PSCAD/EMTDC herein. Taking $Y_{dd}$ as an example (and the consideration of variation of $I_d$ is illustrated herein), Fig. 7 shows errors of the real and imaginary parts of $Y_{dd}$. It can be seen that the errors between the generated admittances and the measured admittances are very small, which verifies that the obtained admittance identification model has good accuracy.

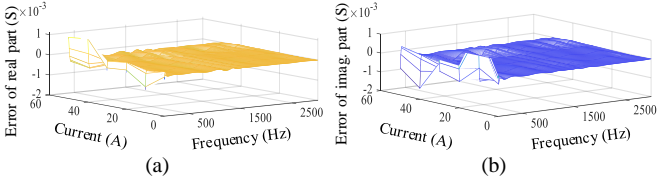

Fig. 8. Errors of (a) real and (b) imaginary parts of $Y_{dd}$ of inverter.

## III. SELF-ADAPTION STRATEGY OF ACTIVE DAMPER FOR SYSTEM'S STABILE MARGIN IMPROVEMENT

### A. Stability Assessment of Inverter-based System

Fig. 9 depicts impedance-based equivalent circuit of multiple paralleled GFL inverter-based system with SAD in dq frame, in which the GFL inverter is equivalent to a current source $I_{S,i}$ ($i=1, 2, \ldots, n$) paralleled with admittance $Y_{eq,i}$; the SAD is also equivalent to a current source $I_{S,ad}$ paralleled with admittance $Y_{ad}$; and the main grid is equivalent to a voltage source $V_g$ series with the impedance $Z_g$.

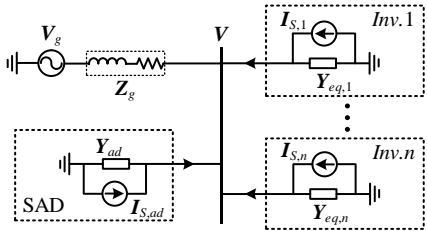

Fig. 9. Impedance-based equivalent circuit of multiple paralleled GFL inverter-based system with SAD in dq frame.

Firstly, according to circuit theory, in the absence of the SAD, the relationship among the current sources $I_{S,i}$, grid voltage $V_g$ and PCC voltage $V$ can be expressed as

$$\underbrace{\left(\sum_{i=1}^{n} I_{S,i} + Z_g^{-1} V_g \right)}_{I_{SS}} = \underbrace{\left(\sum_{i=1}^{n} Y_{eq,i} + Z_g^{-1}\right)}_{Y_{PCC}} \cdot V = \left(I + \underbrace{\sum_{i=1}^{n} Y_{eq,i} \cdot Z_g}_{L_m}\right) \cdot V \quad (6)$$

where $I$ is a 2×2 unit matrix, $I_{SS}$ is the equivalent current source of overall system, $Y_{PCC}$ is the nodal admittance matrix of PCC, and $L_m$ is the return-ratio matrix of the system.

Then the PCC voltage $V$ can be calculated as

$$V = Y_{PCC}^{-1} \cdot I_{SS} = \left(I + L_m\right)^{-1} \cdot I_{SS} \quad (7)$$

According to the impedance-based stability criteria [12], the system's stability can be assessed either by the return-ratio matrix $L_m$ based on GNC [22], or by the nodal admittance matrix $Y_{PCC}$ based on the impedance-sum-type criterion (derived from Cauchy's theorem) [23]. For the latter, given that each inverter is self-stable (which is the basic requirement for the inverter manufacturers), the system's stability is determined by whether the Nyquist trajectory of eigenvalues of $Y_{PCC}$ encircle point (0, 0) or not.

The eigenvalue Nyquist trajectory-based stability criterion can also be equivalent to the form of trajectories of the eigenvalue's real and imaginary parts as functions of frequency [20], [24]. As illustrated in Fig. 10 (a), assume that the Nyquist plot of eigenvalue $\lambda$ encircles point (0, 0), and it crosses the Real Axis (the intersection marked as point A) at crossing frequency $f_{cr}$. Then, plotting the real and imaginary parts of $\lambda$, respectively, as functions of frequency in real-imaginary plots. It can be seen that Im[$\lambda$]=0 and Re[$\lambda$]<0 at $f_{cr}$. In contrast, as shown in Fig. 10 (b), the Nyquist plot of eigenvalue $\lambda$ does not encircle point (0, 0) when the system is stable, then at the frequency of Im[$\lambda$] equal to 0, Re[$\lambda$] is positive, i.e., Im[$\lambda$]($f_{cr}$)=0 and Re[$\lambda$]($f_{cr}$)>0. This indicates a clue for the system's stability enhancement—compensating the real part of $\lambda$ at its crossing frequency $f_{cr}$ to make Re[$\lambda$]($f_{cr}$)>0, which will provide theoretical guidance and further discussion for the adaptive active damping method in the following text. It should be noted that for the conditionally stable system [25], the trajectory-based stability criterion in a real-imaginary plot will be somewhat complicated, which can refer to [20].

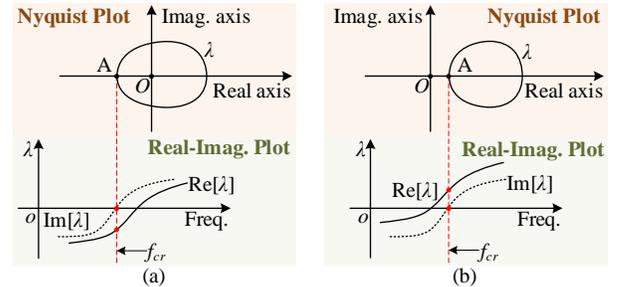

Fig. 10. Nyquist plot of eigenvalue and its real and imaginary parts as functions of frequency: (a) unstable, and (b) stable.

In this paper, a system with two GFL inverters is used as the study case, whose gird voltage is 380V/50Hz, and the inverters' parameters can be found in Table II. Fig. 11 shows the trajectories of eigenvalues of $Y_{PCC}$ in a weak grid, where $L_g$=4 mH and $R_g$=0.2 Ω. The Nyquist plots show that eigenvalue $\lambda_2$ encircles point (0, 0); meanwhile, the real-imaginary plots indicate that at the crossing frequency $f_{cr}$ of $\lambda_2$, Im[$\lambda_2$]=0 and Re[$\lambda_2$]<0, which means the system is unstable.

TABLE II
PARAMETERS OF GFL INVERTERS OF CASE SYSTEM

| Parameters | Inv. 1 | Inv. 2 |
|---|---|---|
| DC-link voltage $V_{dc}$ (V) | 800 | 800 |
| Inductance of filter inductor $L$ (mH) | 3 | 2.5 |
| Resistance of filter resistor $R_L$ (Ω) | 0.015 | 0.01 |
| D channel current $I_d$ (A) | 50 | 60 |
| Q channel current $I_q$ (A) | 0 | 0 |
| Proportional gain of current controller $k_{pi}$ | 18 | 15 |
| Integral gain of current controller $k_{ii}$ (s$^{-1}$) | 300 | 300 |
| Proportional gain of PLL $k_{pPLL}$ | 5 | 5 |
| Integral gain of PLL $k_{iPLL}$ (s$^{-1}$) | 100 | 100 |
| Sampling frequency $f_s$ (kHz) | 10 | 10 |

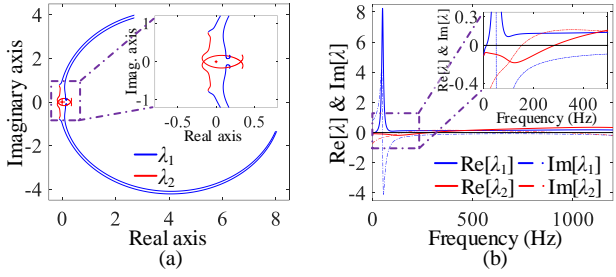

Fig. 11. Trajectories of eigenvalues of $Y_{PCC}$ with $L_g$=4 mH, $R_g$=0.2 Ω: (a) Nyquist plot, and (b) real-imaginary plot.

In this paper, the eigenvalue with the minimum real part at its crossing frequency is defined as the critical eigenvalue, which is the major concern for system's stability enhancement. Fig. 12 depicts the variations of the critical eigenvalue with the increase of inverters' output power or grid impedance, in which the parameters are increasing in the direction of arrows. It can be seen that with the increase of the inverters' output power or grid impedance, the imaginary part of the critical eigenvalue increases and its crossing frequency decreases; meanwhile, the real part of the critical eigenvalue decreases at its crossing frequency, thus the system's stability deteriorates.

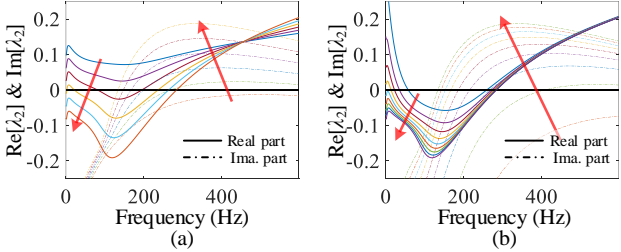

Fig. 12. Real-imaginary plots of critical eigenvalue of $Y_{PCC}$ when system's parameters varying: (a) output power of inverters increasing from 0 to 100%, and (b) grid impedance increasing from 1 mH+0.05 Ω to 5 mH+0.25 Ω.

The SAD for stability enhancement is supposed to compensate the real part of the critical eigenvalue at its crossing frequency so that the system has a sufficient stable margin. As illustrated in Fig. 13, in terms of stability enhancement for the case system, its critical eigenvalue is supposed to be compensated in two aspects: one is to make Re[$\lambda_2$] increased at $f_{cr}$, and the other is to make Im[$\lambda_2$] decreased so $f_{cr}$ becomes increased thereby Re[$\lambda_2$]($f_{cr}$) increased.

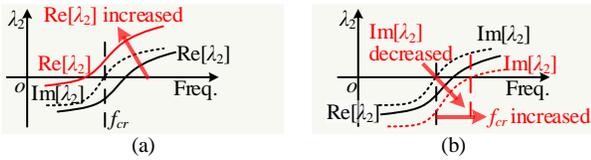

Fig. 13. Conditions of enhancement for the system's stability: (a) Re[$\lambda$] increased, and (b) Im[$\lambda$] decreased.

### B. Control Strategy for SAD and Its Damping Compensation Mechanism for Stability Enhancement

*1) Control strategy and parameter design for SAD:* Fig. 14 shows the control strategy of damping for SAD. In the proposed strategy, a second-order bandpass filter (SBPF) is used to extract the target components of PCC voltage in the dq frame, and what is more, a lag compensator (LAC) [25] is applied to modify the phase shift resulting by the SBPF, then multiplied by a coefficient -$H_v$ to get $v_{av}$, $v_{bv}$, and $v_{cv}$ for sinusoidal pulse width modulation (SPWM).

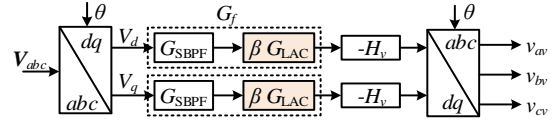

Fig. 14. Control strategy of damping for SAD.

The SBPF can effectively suppress the components out of its passband range, and its frequency domain characteristic is determined by two critical parameters, i.e., the center frequency and bandwidth, as illustrated in Fig. 15. Compared with the second-order high pass filter used in [20] for extracting the non-fundamental components of PCC voltage, the SBPF will not bring high-frequency disturbances of the grid in the control system. Herein, the transfer function of SBPF is

$$G_{\text{SBPF}} = \frac{2\pi Bs}{s^2 + 2\pi Bs + \omega_c^2} \quad (8)$$

where $\omega_c$ and $B$ denote the center frequency and bandwidth of SBPF, respectively.

The transfer function of LAC is

$$G_{\text{LAC}} = \frac{\tau s + 1}{\beta \tau s + 1} \quad (9)$$

As shown in Fig. 15, $\omega_1$ and $\omega_2$ are LAC's two cut-off frequencies, where $\omega_1$=1/$\beta\tau$, and $\omega_2$=1/$\tau$. LAC can lead to the maximum phase $\varphi_m$ at the frequency $\omega_m$, where $\varphi_m$ and $\omega_m$ are given as following

$$\omega_m = 1/\sqrt{\beta}\tau, \quad \varphi_m = \arctan\left[(\beta-1)/(2\sqrt{\beta})\right] \quad (10)$$

Eq. (10) means that $\varphi_m$ is determined by $\beta$, and $\omega_m$ is determined by both $\beta$ and $\tau$. A small $\beta$ will provide insufficient $\varphi_m$, yet too large $\beta$ will degrade the performance of $G_f$ on suppressing the low-frequency disturbances. In this paper the value of $\beta$ is designed to be 20, then it can get that $\varphi_m$=64.8° by (10).

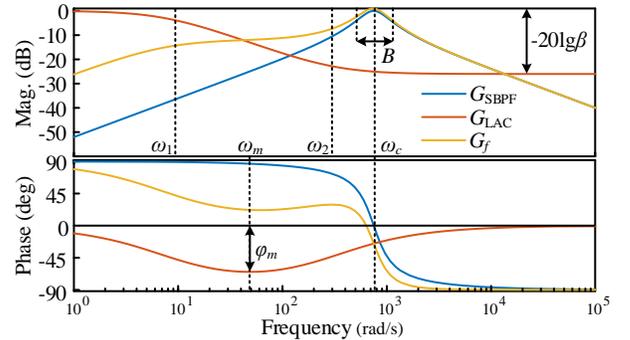

Fig. 15. Bode plots of transfer functions of $G_{\text{SBPF}}$, $G_{\text{LAC}}$ and $G_f$.

In this paper, the center frequency $\omega_c$ of SBPF is selected as one of the parameters for self-adaptive control, which means that as the operating points of the system change, $\omega_c$ will change accordingly. Recall that the LAC is applied to modify the phase shifting led by the SBPF, thus the characteristic frequencies of LAC, i.e., $\omega_1$, $\omega_2$, and $\omega_m$, are also supposed to vary accordingly. In this paper, $\beta$ and $\tau$ are designed to be 20 and 2.5/$\omega_c$, respectively. As shown in Fig. 15, the amplitude-frequency characteristic of $G_f$ is very close to that of $G_{\text{SBPF}}$ in the frequency range above $\omega_2$, which has the ability to suppress high-frequency disturbances. In the frequency range below $\omega_2$, compared to $G_{\text{SBPF}}$, $G_f$ still has the ability of suppressing low-frequency disturbances to a certain extent; what is important is that the phase of $G_f$ is compensated by $G_{LAC}$, the purpose of which will be explained later (see Fig. 17).

2) *Output admittance of SAD*: Essentially, the active dampers improve the stable margin of the inverter-based systems by changing the nodal admittance of the network where they are integrated. Thus the output admittance of SAD should be properly designed. Fig. 16 illustrates the block diagram of the current control loop of SAD, where the outer dc-link voltage control loop and PLL have been neglected considering that the dynamic response of the inner current control loop is normally faster than those of the outer control loops. According to Fig. 15, the output admittance of SAD is

$$Y_{ad} = -\frac{V_{dq}}{I_{dq}} = \frac{1+H_v G_f G_d}{Z_f + G_i G_d} \quad (11)$$

where $Z_f = sL_f + R_f$, $L_f$ and $R_f$ denote the inductance and resistance of filter, respectively; $G_i = k_{cp} + k_{ci}/s$, $k_{cp}$ and $k_{ci}$ denote proportional gain and integral gain of current proportional-integral (PI) controller, respectively; $G_d$ is the time delay introduced by the digital control, which is expressed as

$$G_d = e^{-sT_d} \approx \frac{1 - 0.5T_d s + 1/12 T_d^2 s^2}{1 + 0.5T_d s + 1/12 T_d^2 s^2} \quad (12)$$

where $T_d$ is the delay time, $T_d = 1.5/f_s$, and $f_s$ is the sampling frequency.

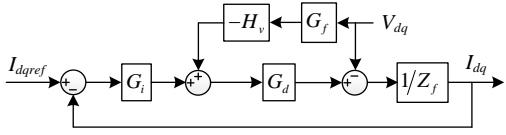

Fig. 16. Block diagram of the current control loop of SAD.

Then, ignoring the coupling effect between *d*-axis and *q*-axis, the approximation of the SAD's output admittance matrix can be expressed as $Y_{eq} \approx [Y_{ad}, 0; 0, Y_{ad}]$.

3) *Damping compensation mechanism of SAD for Stability Enhancement*: According to the eigenvalue decomposition theory [26], the relationship between the nodal admittance matrix $Y_{PCC}$ and its eigenvalue $\lambda$ satisfies

$$|\lambda I - Y_{PCC}| = 0 \quad (13)$$

When the SAD is shunted into PCC, the nodal admittance matrix of PCC becomes $Y_{PCC} + Y_{eq}$, then its eigenvalue $\lambda_s$ satisfies

$$|\lambda_s I - (Y_{PCC} + Y_{eq})| = 0 \quad (14)$$

Substituting the output admittance $Y_{eq}$ into (15) yields

$$|\lambda_s I - (Y_{PCC} + Y_{ad}I)| = |(\lambda_s - Y_{ad})I - Y_{PCC}| = 0 \quad (15)$$

According to (13) and (15), it can be obtained that $\lambda_s = \lambda + Y_{ad}$, which means that the effect of the shunted SAD on the system's stability can be evaluated by the superposition of admittance $Y_{ad}$ and the critical eigenvalue of the original system without SAD.

Fig. 17 shows the real and imaginary parts of $Y_{ad}$. It indicates that the imaginary part of $Y_{ad}$ with LAC has a smaller value than that of $Y_{ad}$ without LAC in the range below their crossing frequencies; while the real part of $Y_{ad}$ with LAC has a larger value than that of $Y_{ad}$ without LAC in the range over their crossing frequencies. Recall that for the case system, the imaginary part of its critical eigenvalue is supposed to be compensated to decrease while its real part is supposed to be compensated to increase. Therefore, the proposed SAD with LAC has better damping compensation performance on stability enhancement compared with the case without LAC, which demonstrates the modifying effect of LAC on the admittance of SAD.

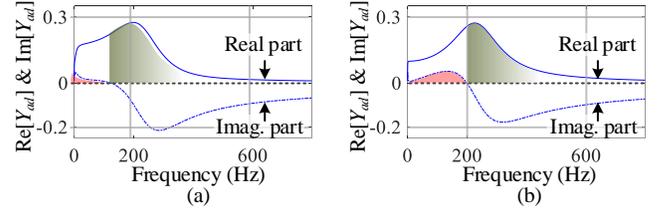

Fig. 17. Real and imaginary parts of output admittance of SAD: (a) modified by LAC, and (b) without LAC.

Fig. 18 demonstrates the eigenvalues of $Y_{PCC}$ before (the blue lines) and after (the bold green lines) compensated with $Y_{ad}$ of SAD (the red lines), where the parameters of SAD are listed in Table II. Fig. 18 (a) indicates that after being compensated with $Y_{ad}$, the real part of critical eigenvalue of $Y_{PCC}$ is improved at its new crossing frequency $f_{cr2}$, hence the system becomes stable. In Fig. 18 (b), the other eigenvalue of $Y_{PCC}$ shows no deterioration of stability after being compensated with $Y_{ad}$.

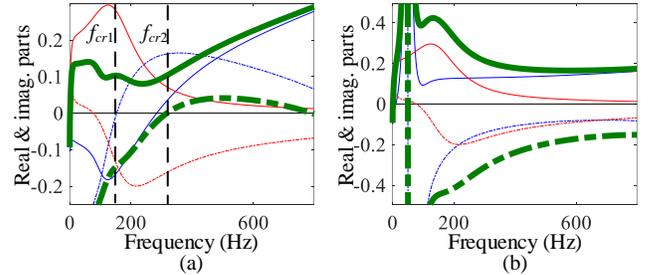

Fig. 18. Eigenvalues of $Y_{PCC}$ before and after compensated with SAD: (a) the critical eigenvalue, and (b) the other eigenvalue.

TABLE III
PARAMETERS OF SAD

| Parameters | Values |
| --- | --- |
| DC-link voltage $V_{dc}$ (V) | 750 |
| DC-link Capacitance $C_{dc}$ (μF) | 5000 |
| Inductance of filter $L_f$ (mH) | 3 |
| Resistance of filter $R_f$ (Ω) | 0.01 |
| Proportional gain of voltage controller $k_{cp}$ | 0.5 |
| Integral gain of voltage controller $k_{ci}$ (s$^{-1}$) | 5 |
| Feedback coefficient of PCC voltage $H_v$ | 2 |
| Proportional gain of current controller $k_{cp}$ | 10 |
| Integral gain of current controller $k_{ci}$ (s$^{-1}$) | 20 |
| Proportional gain of PLL $k_{pPLL}$ | 0.5 |
| Integral gain of PLL $k_{iPLL}$ (s$^{-1}$) | 50 |
| Bandwidth of SBPF (Hz) | 200 |
| Coefficient $\beta$ of LAC | 20 |
| Sampling frequency $f_s$ (kHz) | 20 |

Fig. 19 shows the simulation results of grid currents of the case system. The parameters of inverters and SAD can be found in Table II and Table III; besides, the gird impedance is set as $L_g$=4 mH and $R_g$=0.2 Ω. It can be seen that, with SAD (where $\omega_c$=1005.31 rad/s and $H_v$=1.8), the system is stable when inverters' output currents ramp up from half to full of the rated value; however, at 0.38 s the SAD is removed and the system becomes unstable. The simulation results are consistent with the theoretical analysis, which also underlines the necessity of an active damper for the system's stability.

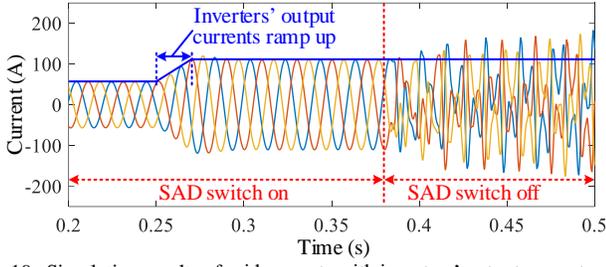

Fig. 19. Simulation results of grid currents with inverters' output currents ramp up.

### C. Self-Adaptive Strategy Based on ANN for System's Stability Enhancement

The damping compensation mechanism of SAD is analyzed above, and it reveals that the system's critical eigenvalue can be effectively compensated by a properly designed $Y_{ad}$ of SAD for stability enhancement. On the other hand, as illustrated in Fig. 12, considering that the system's parameters (i.e., inverters' output power and grid impedance) vary during operation, the system's critical eigenvalue of $Y_{PCC}$ will change. Therefore, the $Y_{ad}$ of SAD is supposed to be also changed adaptively in accordance with the system's parameters, and to ensure the system has a sufficient stability margin, thereby eliminating the potential instability risk of the system.

Fig. 20 shows the clusters of the real part (the solid lines) and imaginary part (the dotted lines) of the output admittance of SAD with parameters (i.e., $\omega_c$ and $H_v$) varying. It can be found that as $\omega_c$ increases, the frequency of the peak value of Re[$Y_{ad}$] also increases, which means its optimum compensating area shifts to a higher frequency range. Recall that, as illustrated in Fig. 12, the crossing frequency of critical eigenvalue varies as the system's parameters change, thus $\omega_c$ can be designed to be changeable for self-adaptive active damping compensation. On the other hand, Fig. 20 shows that the real and imaginary parts of $Y_{ad}$ are in proportion to $H_v$, which means that $H_v$ can be selected as a regulatory factor to adjust the degree of compensation of SAD.

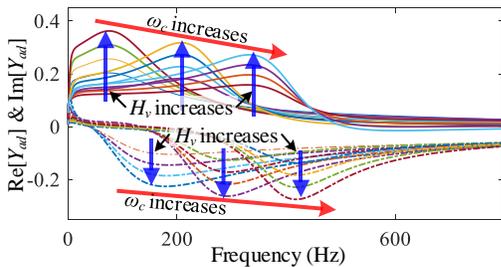

Fig. 20. Clusters of real and imaginary parts of output admittance of SAD with parameters varying.

The impact of parameter variations on the system's critical eigenvalue—as well as the requirement of $Y_{ad}$ with appropriate parameters for stability enhancement—follows a certain pattern. However, due to the non-linear relationship between the system's critical eigenvalue or $Y_{ad}$ of SAD and their related parameters, it is challenging to directly establish a quantitative calculation relationship between the system's parameters and SAD's parameters from the perspective of the system's stability enhancement. To address this problem, this paper uses the ANN-based technique to establish the mapping between the system's parameter (i.e., inverters' output power and grid impedance) variations and SAD's parameters (i.e., $\omega_c$ and $H_v$), which can enable $Y_{ad}$ to match the system's critical eigenvalue and make the system have sufficient stable margin.

The ANN structure for matching $Y_{ad}$ to the system's critical eigenvalue also consists of input layer, hidden layer, and output layer. The inverters' output d-axis currents (related to active power) and the grid impedance are chosen to be included in the input layer, and the output layer contains the parameters $\omega_c$ and $H_v$ of SAD. Besides, 1 hidden layer with 10 neurons is used in the training of the ANN model. Note that these inverters normally operate in unity power factor mode, their output q-axis currents that relate to reactive power are not considered.

The next important procedure is to acquire the dataset for the model training. As illustrated in Fig. 21, firstly, the inverters' output currents and PCC voltage are fed to the admittance model generated in Section II to get inverters' admittances; then together with $Z_g$, the system's stability can be assessed and its critical eigenvalue can be identified. Finally, the parameters of SAD can be designed by trial and error based on the damping compensation mechanism discussed in Section III-B. It should be noted that to make the system have a sufficient stable margin, the real part of the critical eigenvalue at its crossing frequency should be greater than a threshold value $\sigma_{thd}$, i.e., Re[$\lambda$]($f_{cr}$)≥$\sigma_{thd}$. A larger $\sigma_{thd}$ (involving larger $H_v$) makes the system have a larger stable margin, but it also means a larger port current will flow into the SAD and lead to increased power loss [21]. In this paper, $\sigma_{thd}$ is set as 0.1.

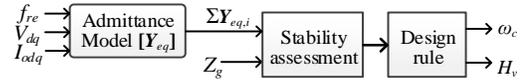

Fig. 21. Off-line procedure to acquire dataset for model training of SAD's parameters.

The system's multiple operating points are composed of the permutation and combination of different inverters' output currents and grid impedance, which are set as: $I_{d1}$=[0 A, 12.5 A, 25 A, 37.5 A, 50 A]$^T$, $I_{d2}$=[0 A, 15 A, 30 A, 45 A, 60 A]$^T$, and $Z_g$=[1.5 mH + 0.075 Ω, 2.5 mH + 0.125 Ω, 3.5 mH + 0.175 Ω, 4.5 mH + 0.2275 Ω, 5.5 mH + 0.275 Ω]$^T$. The parameters of SAD are properly designed under these operating points by trial and error based on the damping compensation mechanism. Then the dataset for the model training of SAD's parameters can be acquired. Fig. 22 shows the training, validation, and test performance of the model training, which indicates that the training is effective.

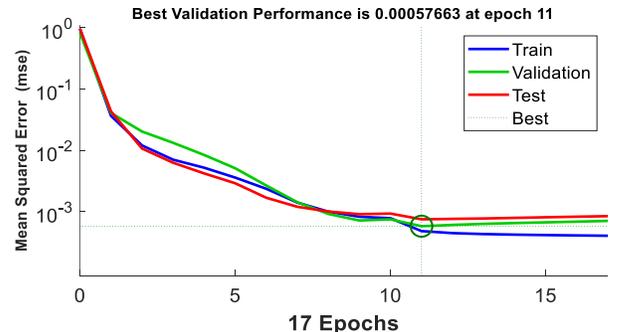

Fig. 22. Training, validation, and test performance of the model training of SAD's parameters.

## IV. EXPERIMENTAL VERIFICATION

To verify the proposed grid impedance estimation method and the self-adaptive active damping method based on SAD for stability enhancement, experiments of the case system are carried out on the platform of on the platform of control hardware in the loop (CHIL). Parameters of inverters and SAD given in Tables I and II are used in experiments.

Fig. 23 shows the experimental results of grid impedance estimation, where two scenarios (i.e., $L_g$=3 mH, $R_g$=0.15 Ω, and $L_g$=4 mH, $R_g$=0.2 Ω) are taken as examples. It can be seen that the proposed method can accurately estimate the grid impedance within tens of milliseconds.

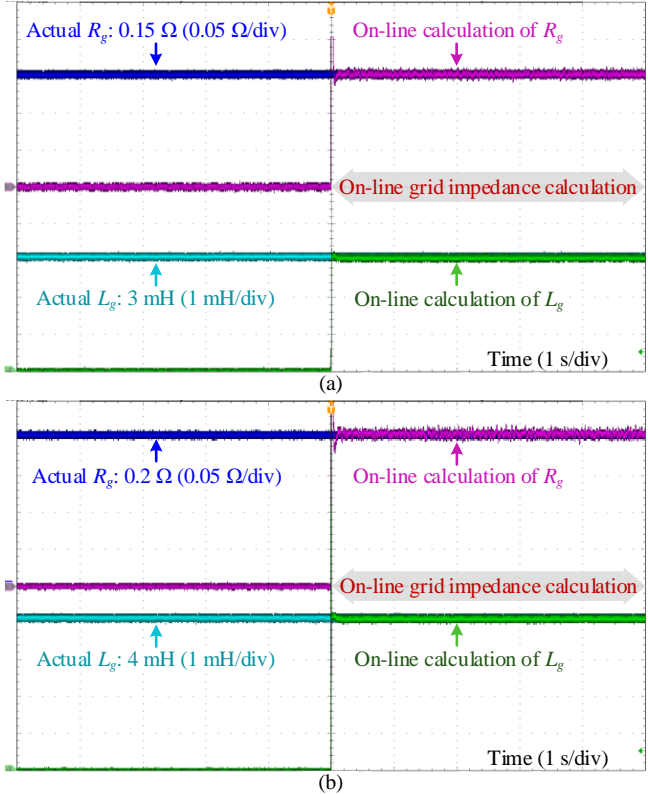

Fig. 23. Grid impedance estimation with: (a) $L_g$=3 mH and $R_g$=0.15 Ω, and (b) $L_g$=4 mH and $R_g$=0.2 Ω.

Fig. 24 shows the experimental results of grid currents under different system's operating points. It can be seen that with SAD, the grid currents can be stable when inverters' currents ramp up or down. As shown in Fig. 24 (a), when the inverters' output power is 0.5 p.u, the SAD is operating with $\omega_c$ = 1315.26 rad/s and $H_v$ = 0.42. As the inverter parameters change, $\omega_c$ and $H_v$ also adaptively adjust according to the stability margin requirements. When the inverter output power becomes 1.0 p.u, $\omega_c$ and $H_v$ are adaptively adjusted to 1386.01 rad/s and 1.83, and the system operates stably. However, when SAD switched off, the system became unstable.

Similarly, Fig. 24 (b) shows the experimental results of inverter output power reduction with SAD. When the inverters' output power is 1.0 p.u, the SAD is operating with $\omega_c$ = 1206.37 rad/s and $H_v$ = 1.39. When the inverter output power becomes 0.75 p.u, $\omega_c$ and $H_v$ are adaptively adjusted to 141.94 rad/s and 0.69, and the system operates stably. However, when SAD switched off, the system became unstable. The experimental results shown in Fig. 24 verify the effectiveness of the proposed self-adaptive active damping method.

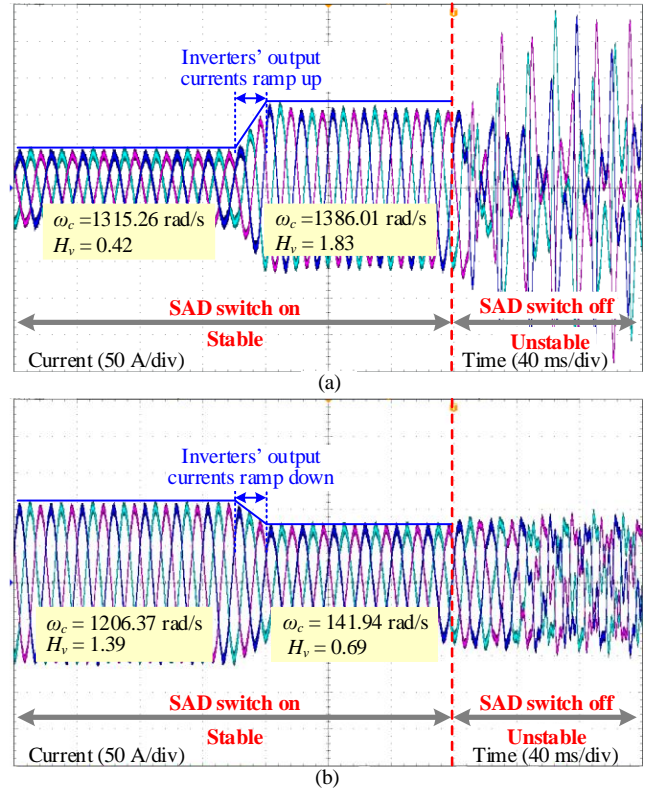

Fig. 24. Grid currents with inverters' output currents: (a) ramp up from 50% to 100%, where $L_g$=4 mH and $R_g$=0.2 Ω, and (b) ramp down from 100% to 75%, where $L_g$=3 mH and $R_g$=0.15 Ω.

## V. CONCLUSIONS

This paper proposes a self-adaptive active damping method for the stability enhancement of inverter-based systems, which is a promising solution to eliminate the risk of potential instability for systems with black-box inverters. Considering the variations of operating points, a widely-applicable and engineering-friendly method is presented to on-line calculate the grid impedance. What is more, a self-adaptive active damper as well as its control strategy for self-adaptive active damping is proposed, which can realize on-line parameter-tuning for SAD to compensate the system's damping according to operating points. Besides, the SAD's damping compensation mechanism for the system's stability enhancement is investigated and revealed, based on which the parameters of active dampers can be properly designed under different operating points.

## REFERENCES

[1] H. Zhang, G. Xiao, Z. Liu and Y. Zhou, "Impedance-Based Stability Analysis and On-Site Stability Evaluation of Three-Phase Active Voltage Conditioner Embedded System," *IEEE Trans. Power Electron.*, vol. 38, no. 12, pp. 16061-16071, Dec. 2023.
[2] X. Wang and F. Blaabjerg, "Harmonic stability in power electronic-based power systems: Concept, modeling, and analysis," *IEEE Trans. Smart Grid.*, vol. 10, no. 3, pp. 2858–2870, May 2019.
[3] B. Wen, D. Boroyevich, R. Burgos, P. Mattavelli, and Z. Shen, "Analysis of D-Q small-signal impedance of grid-tied inverters," *IEEE Trans. Power Electron.*, vol. 31, no. 1, pp. 675–687, Jan. 2016.
[4] Wei Liu, Xiaorong Xie, Jan Shair, Jiaqi Zhang, "Stability Region Analysis of Grid-Tied Voltage Sourced Converters Using Variable Operating Point

Impedance Model", *IEEE Trans. Power Electron.*, vol.38, no.2, pp.1125-1137, 2023.

[5] J. Z. Zhou, H. Ding, S. Fan, Y. Zhang and A. M. Gole, "Impact of Short-Circuit Ratio and Phase-Locked-Loop Parameters on the Small-Signal Behavior of a VSC-HVDC Converter," *IEEE Trans. Power Del.*, vol. 29, no. 5, pp. 2287-2296, Oct. 2014.

[6] N. Hoffmann and F. W. Fuchs, "Minimal Invasive Equivalent Grid Impedance Estimation in Inductive–Resistive Power Networks Using Extended Kalman Filter," *IEEE Trans. Power Electron.*, vol. 29, no. 2, pp. 631-641, Feb. 2014.

[7] N. Mohammed, T. Kerekes and M. Ciobotaru, "An Online Event-Based Grid Impedance Estimation Technique Using Grid-Connected Inverters," *IEEE Trans. Power Electron.*, vol. 36, no. 5, pp. 6106-6117, May 2021.

[8] J. H. Cho, K. Y. Choi, Y. W. Kim and R. Y. Kim, "A novel P-Q variations method using a decoupled injection of reference currents for a precise estimation of grid impedance", in *Proc. IEEE Energy Convers. Congr. Expo.*, pp. 5059-5064, Nov. 2014.

[9] Anastasis Charalambous, Lenos Hadjidemetriou, Marios Polycarpou, "A Sensorless Asymmetric and Harmonic Load Compensation Method by Photovoltaic Inverters Based on Event-Triggered Impedance Estimation", *IEEE Trans. Ind. Electron.*, vol.70, no.10, pp.10089-10100, 2023.

[10] Jingrong Yu, Wenjing Liu, Jianwen Sun, Fu Zhang, Yuxiang Yang, "An improved grid impedance estimator for grid-forming converters in consideration of controller dynamics", *International Journal of Electrical Power & Energy Systems*, vol.154, pp.109424, 2023.

[11] J. Sun, "Impedance-Based stability criterion for grid-connected inverters," *IEEE Trans. Power Electron.*, vol. 26, no. 11, pp. 3075–3078, Nov. 2011.

[12] W. Cao, Y. Ma, L. Yang, F. Wang and L. M. Tolbert, "D–Q Impedance Based Stability Analysis and Parameter Design of Three-Phase Inverter-Based AC Power Systems," *IEEE Trans. Ind. Electron.*, vol. 64, no. 7, pp. 6017-6028, July 2017.

[13] A. V. Timbus, P. Rodriguez, R. Teodorescu and M. Ciobotaru, "Line impedance estimation using active and reactive power variations", in *Proc. IEEE Power Electron. Spec. Conf.*, pp. 1273-1279, 2007.

[14] M. Zhang, X. Wang, D. Yang, et al., "Artificial Neural Network Based Identification of Multi-Operating-Point Impedance Model," *IEEE Trans. Power Electron.*, vol. 36, no. 2, pp. 1231-1235, Feb. 2021.

[15] Y. Li et al., "Machine Learning at the Grid Edge: Data-Driven Impedance Models for Model-Free Inverters," *IEEE Trans. Power Electron.*, vol. 39, no. 8, pp. 10465-10481, Aug. 2024.

[16] Mengfan Zhang, Qianwen Xu, "Deep Neural Network-Based Stability Region Estimation for Grid-Converter Interaction Systems", *IEEE Trans. Ind. Electron.*, vol.71, no.10, pp.12233-12243, 2024.

[17] Chen Zhang, Nenad Mijatovic, Xu Cai, Tomislav Dragičević, "Artificial Neural Network-Based Pole-Tracking Method for Online Stabilization Control of Grid-Tied VSC", *IEEE Trans. Ind. Electron.*, vol.69, no.12, pp.13902-13909, 2022.

[18] C. Zhang, M. M. Mardani and T. Dragičević, "Adaptive Multi-Parameter-Tuning for Online Stabilization Control of Grid-Tied VSC: An Artificial Neural Network-Based Method," *IEEE Trans. Power Del.*, vol. 37, no. 4, pp. 3428-3431, Aug. 2022.

[19] A. Adib, F. Fateh and B. Mirafzal, "Smart Inverter Stability Enhancement in Weak Grids Using Adaptive Virtual-Inductance," *IEEE Trans. Power Electron.*, vol. 57, no. 1, pp. 814-823, Jan.-Feb. 2021.

[20] Y. Li, X. Wu, Z. Shuai, Q. Zhou, H. Chen and Z. J. Shen, "A Systematic Stability Enhancement Method for Microgrids With Unknown-Parameter Inverters," *IEEE Trans. Power Electron.*, vol. 38, no. 3, pp. 3029-3043, March 2023.

[21] Z. Lin, X. Ruan, H. Zhang and L. Wu, "A Hybrid-Frame Control Based Impedance Shaping Method to Extend the Effective Damping Frequency Range of the Three-Phase Adaptive Active Damper," *IEEE Trans. Ind. Electron.*, vol. 70, no. 1, pp. 509-521, Jan. 2023.

[22] W. Cao, Y. Ma and F. Wang, "Sequence-Impedance-Based Harmonic Stability Analysis and Controller Parameter Design of Three-Phase Inverter-Based Multibus AC Power Systems," *IEEE Trans. Power Electron.*, vol. 32, no. 10, pp. 7674-7693, Oct. 2017.

[23] F. Liu, J. Liu, H. Zhang and D. Xue, "Stability Issues of Z + Z Type Cascade System in Hybrid Energy Storage System (HESS)," *IEEE Trans. Power Electron.*, vol. 29, no. 11, pp. 5846-5859, Nov. 2014.

[24] Y. Li, Z. Shuai, J. Fang and X. Wu, "Stability improvement method for multi-converter distribution systems based on impedance sensitivity," *CSEE J. Power Energy Syst.*, doi: 10.17775/CSEEJPES.2021.04160.

[25] K Ogata, Modern Control Engineering, 5th ed. New Jersey, USA: Prentice Hall, 2010.

[26] Leon S J, " Linear Algebra with Application," 8th ed, New York: Macmillan, 2010.